\documentclass[%
reprint,
superscriptaddress,
twocolumn,
showpacs,
amsmath,amssymb,
aps,
prl,
floatfix,
showkeys,
]{revtex4-2}

\usepackage{graphicx}
\usepackage{dcolumn}
\usepackage{bm}
\usepackage{amsmath}
\usepackage{amssymb}
\usepackage{multirow}
\usepackage{color}
\usepackage[T1]{fontenc}
\usepackage[colorlinks=true,linkcolor=blue,citecolor=blue]{hyperref}
\usepackage{booktabs}
\usepackage{tabularx}
\usepackage[english]{babel}
\usepackage[utf8]{inputenc}
\usepackage{enumerate}
\usepackage[cmyk,dvipsnames]{xcolor}

\usepackage[normalem]{ulem}

\definecolor{pacificb}{HTML}{1CA9C9}

\date{ \today}

\begin{document}

\title{Enhancing thermal stability of optimal magnetization reversal in nanoparticles}

\author{Mohammad H. A. Badarneh}
\affiliation{Science Institute of the University of Iceland, 107 Reykjav\'ik, Iceland}

\author{Grzegorz J. Kwiatkowski}
\affiliation{Science Institute of the University of Iceland, 107 Reykjav\'ik, Iceland}

\author{Pavel F. Bessarab}
\email[Corresponding author: ]{pavel.bessarab@lnu.se}
\affiliation{Science Institute of the University of Iceland, 107 Reykjav\'ik, Iceland}
\affiliation{Department of Physics and Electrical Engineering, Linnaeus University, SE-39231 Kalmar, Sweden}

\begin{abstract}
Energy-efficient switching of nanoscale magnets requires the application of a time-varying magnetic field characterized by microwave frequency. 
At finite temperatures, even weak thermal fluctuations create perturbations in the magnetization that can accumulate in time, break the phase locking between the magnetization and the applied field, and eventually compromise magnetization switching. 
It is demonstrated here 
that the magnetization reversal is mostly disturbed by unstable perturbations arising in a certain domain of the configuration space of a nanomagnet. 
The instabilities can be suppressed and the probability of magnetization switching enhanced by applying an additional stimulus such as a  weak longitudinal magnetic field that ensures bounded dynamics of the perturbations. 
Application of the stabilizing longitudinal field to a uniaxial nanomagnet makes it possible to reach a desired probability of magnetization switching even at elevated temperatures. The principle of suppressing instabilities provides a general approach to enhancing thermal stability of magnetization dynamics.
\end{abstract}

\maketitle
\textit{Introduction ---} 
Identification of energy-efficient methods for controlling magnetization is both fundamentally interesting and technologically relevant, e.g., in the development of magnetic memory devices. While magnetization switching in magnetic recording is conventionally achieved by applying a static external magnetic field opposite to the initial magnetization direction, previous studies have demonstrated that the energy cost of this process can be reduced by applying time-varying stimuli, such as a microwave magnetic field~\cite{thirion2003,sun2006,rivkin2006,woltersdorf2007,cai2013reversal}. For a uniaxial monodomain particle, the optimal magnetization reversal is achieved by a rotating magnetic field synchronized with the precessional dynamics of the magnetic moment~\cite{sun2006a,barros2011,barros2013,kwiatkowski2021optimal}. 

The assessment of the stability of energy-efficient switching protocols with respect to ever-present thermal fluctuations is an important problem. The thermal fluctuations perturb the phase locking between the magnetization and the external stimulus. As a result, the magnetization switching can be compromised unless the energy barrier between the initial and final states is much larger than the thermal energy, and the switching time does not exceed a few periods of Larmor precession~\cite{kwiatkowski2021optimal}. This poses a challenge for the realization of energy-efficient switching protocols at elevated temperatures, such as a combination of a microwave and heat-assisted technique. Even at low temperatures, the perturbations in the dynamics can accumulate in time potentially leading to decoherence between the magnetization and the microwave pulse for relatively slow switching which is required for the autoresonance-based protocols~\cite{klughertz2014autoresonant}. In general, the assessment and control of dynamical stability of magnetic systems is a crucial problem~\cite{berkov2002fast, wang2008thermal}.

In this work, we demonstrate that thermal stability of 
magnetization switching in nanoparticles is mostly defined by 
unstable perturbations arising in a certain domain of the configuration space of the system. The instabilities can be suppressed by application of a longitudinal magnetic field, which provides a mechanism for enhancing the thermal stability of optimal magnetization switching induced by rotating magnetic field. We show that the success rate of the switching for a given temperature and switching time can be tuned by adjusting the strength of the stabilizing field. Our results provide a perspective on the control of dynamical stability of magnetic systems subject to thermal fluctuations.  


\begin{figure}[t!]
\centering
\includegraphics[width=0.85\columnwidth]{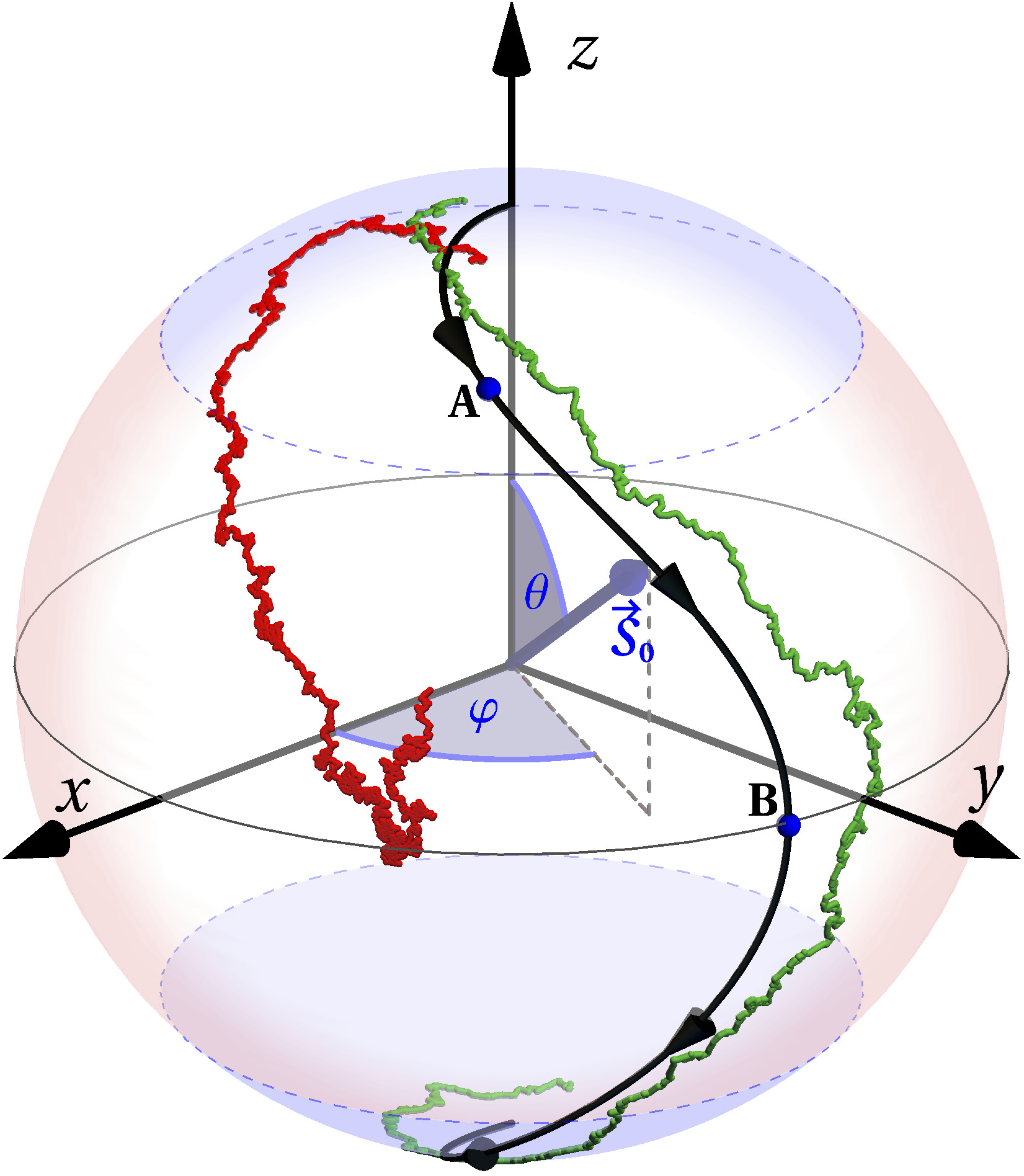}
\caption{\label{fig:system} Calculated dynamics of the magnetic moment for a uniaxial nanoparticle induced by the optimal switching magnetic field. The black line shows the zero-temperature trajectory of the magnetic moment which corresponds to the optimal control path $\vec{s}_0(t)$ for the magnetization switching. The green (red) line shows the trajectory for successful (unsuccessful) switching at finite temperature corresponding to the thermal stability factor $\Delta=20$. Labels A and B show positions of the magnetic moment for which the dynamics of local perturbations in the magnetization is illustrated in the corresponding insets of Fig.~\ref{fig2}. 
The light red (blue) shaded area marks the domain where the perturbation dynamics is unstable (stable). 
The damping factor $\alpha$ is $0.2$, the switching time $T$ is $5\tau_0$. 
}
\end{figure}

\textit{Model and spin dynamics simulations ---} 
We consider energy-efficient magnetization switching 
of a uniaxial monodomain  nanoparticle characterized by normalized magnetic moment $\vec{s}$ and internal energy $ E = -K(\vec{s}\cdot \vec{e}_z)^2/2$, with unit vector $\vec{e}_z$ being the direction and $K>0$ being the strength of the magnetic anisotropy. The switching is induced by an optimal pulse of a rotating magnetic field $\vec{B}_0(t)$ that, for a given switching time, minimizes the energy cost of switching [see Ref.~\cite{kwiatkowski2021optimal} for the exact time dependence of $\vec{B}_0(t)$ as a function of parameters of the nanoparticle]. The switching dynamics is simulated by the time integration of the the Landau-Lifshitz-Gilbert (LLG) equation:
%
\begin{equation} \label{LLG}
(1+\alpha^2)\dot{\vec{s}} = -\gamma \vec{s} \times \vec{B}_\text{eff}  - \alpha \gamma  \vec{s} \times \left(\vec{s} \times \vec{B}_\text{eff} \right),
\end{equation} 
where  the dot denotes the time derivative, $\alpha$ is the Gilbert damping factor, $\gamma$ is the gyromagnetic ratio, and $\vec{B}_\text{eff}\equiv \vec{B}_0+\vec{b}+\vec{\xi}$ is the effective field that, in addition to the switching pulse, includes the internal field $\vec{b}=-\mu^{-1}\partial E/\partial  \vec{s}=\mu^{-1}K(\vec{s}\cdot \vec{e}_z)\vec{e}_z$, with $\mu$ being the magnitude of the magnetic moment, and the stochastic term $\vec{\xi}$ mimicking interaction of the system with the heat bath~\cite{berkov2007}. 
Each simulation involves three stages~\cite{badarneh2023reduction,kwiatkowski2021optimal}: i) Initialization of the magnetic moment close to the energy minimum at $s_z=1$ and equilibration of the system at zero applied magnetic field to establish local Boltzmann distribution at the initial state; ii) Switching where the optimal magnetic field is applied; iii) Final equilibration at zero applied magnetic field. The switching is considered successful if the system is close to the reversed state at $s_z=-1$ at the end of the simulation. 
Proper statistics of switching is obtained by repeating simulations multiple times. 

At zero temperature, the switching trajectory corresponds to optimal control path (OCP) $\vec{s}_0(t)$ between the energy mimima of the system (see the black line in Fig.~\ref{fig:system}): magnetic moment rotates steadily from the initial state at $s_z=1$ to the final state at $s_z=-1$ and simultaneously precesses around the anisotropy axis, where the sense of precession changes at the top of the energy barrier. Magnetization dynamics is synchronized with the switching field so that $\vec{s}_0$ is always perpendicular to $\vec{B}_0$.

At nonzero temperature, thermal fluctuations perturb the magnetization dynamics making the switching trajectory deviate from the OCP (see the green line in Fig.~\ref{fig:system}). The deviation can become so large that the phase locking between the switching pulse and the magnetic moment is lost which may eventually prevent the magnetization reversal (see the red line in Fig.~\ref{fig:system}). 

The success rate of switching depends strongly on the switching time $T$ and the strength of thermal fluctuations, which can be quantitatively described by the thermal stability factor $\Delta$ -- the ratio between the energy barrier separating the stable states and the thermal energy. For $\Delta \gtrsim 70$, which is a standard case for magnetic memory elements~\cite{richter2009,krounbi2015}, and relatively fast switching with $T \lesssim 10 \tau_0$, the switching success rate is close to unity~\cite{kwiatkowski2021optimal}. However, the success rate becomes $0.85$ for $\Delta=20$ and $T=10\tau_0$, and further decreases with decreasing $\Delta$. Furthermore, an increase in the switching time leads to a higher chance for perturbations in the magnetization dynamics to accumulate, thereby increasing the likelihood of unsuccessful switching even for large thermal stability factors. For example, the switching success rate is $0.7$ for $\Delta=70$ and $T=30\tau_0$. These effects make it problematic to realize energy-efficient protocols involving multiple precessions around the anisotropy axis, especially at elevated temperature~\cite{kwiatkowski2021optimal}. In the following, we analyze the local dynamics of perturbations to gain insight into the mechanism of decoherence between the switching pulse and the magnetic moment. This analysis ultimately reveals a method to control the thermal stability of magnetization switching.

\textit{Local dynamics of perturbations ---} The interaction of the nanoparticle with the heat bath results in the 
perturbed trajectory: $\vec{s}(t) = \vec{s}_0(t) + \delta \vec{s}(t)$. 
If the perturbation becomes too large, the coherence between the switching pulse and the magnetic moment will be lost resulting in a failed switching attempt (see red trajectory in Fig. \ref{fig:system}). 
Therefore, the dynamical stability of the system can be investigated by analyzing the time evolution of the perturbation $\delta \vec{s}(t)$. Linearization of Eq. (\ref{LLG}) leads to the following equation of motion for the perturbation: 

\begin{equation} \label{eq_first_order_pert}
   \frac{1+\alpha^2}{\gamma}  \dot{\vec{\epsilon}}(t) =  \left[\begin{array}{cc}
        -\alpha & -1 \\
        1 & -\alpha
    \end{array}\right] \cdot \left[\begin{array}{cc}
        w_1 & 0 \\
        0 & w_2
    \end{array}\right] \cdot \vec{\epsilon}(t).
\end{equation}
Here, $\vec{\epsilon}(t)=(\epsilon_1,\epsilon_2)^T$ is the two-dimensional vector whose components are the coordinates of $\delta \vec{s}$ in the tangent space of $\vec{s}_0(t)$ defined by the eigenvectors of the Hessian of the energy of the system~\cite{varentcova2020}, and $w_1$, $w_2$ are the Hessian's eigenvalues given by the following equations:
\begin{align}
    w_1&=B_r + \frac{K}{\mu}\cos\left(2\theta\right),\\
    w_2&=B_r + \frac{K}{\mu}\cos^2\theta,
\end{align}
where $\theta$ is the polar angle of $\vec{s}_0$ and $B_r$ is the component of the external magnetic field parallel to $\vec{s}_0$. Interestingly, local dynamics of the perturbations does not depend explicitly on the optimal switching pulse, for which $B_r=0$. 

\begin{figure}[t!]
\centering
\includegraphics[width=\columnwidth]{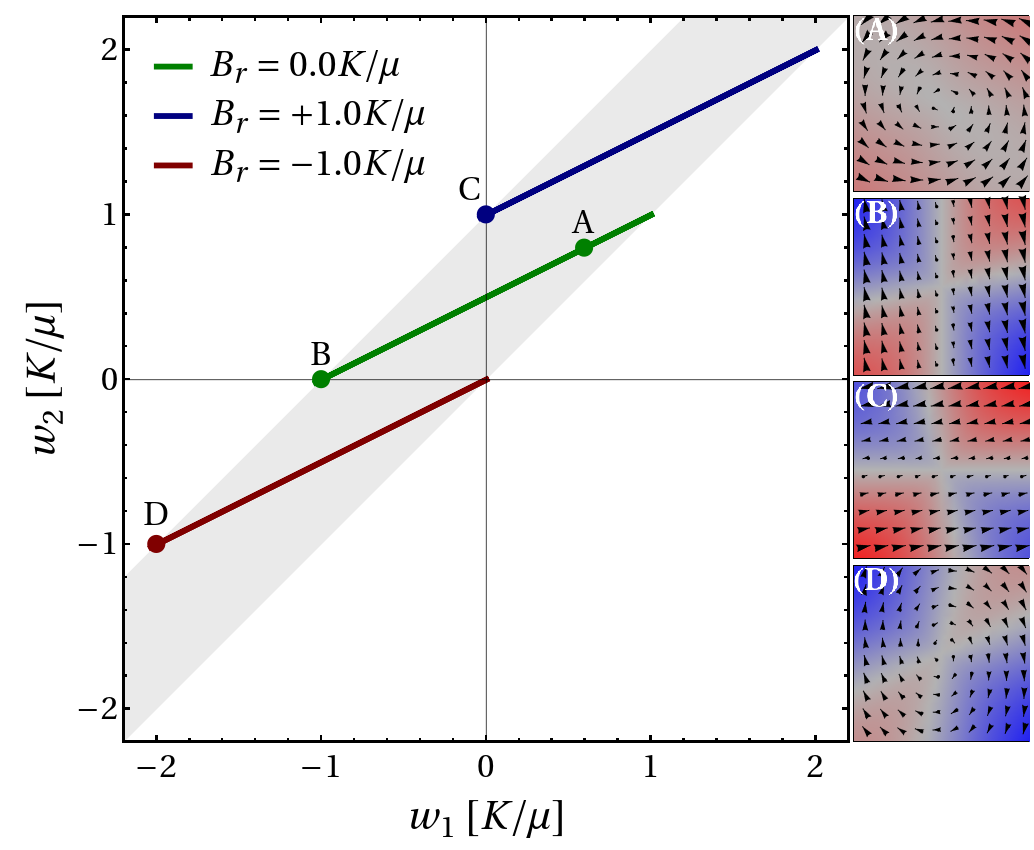}
\caption{\label{fig2} Diagram classifying dynamics of perturbations in the magnetization. 
The green, blue, and red lines show how the Hessian's eigenvalues $w_1$ and $w_2$ 
change along the zero temperature reversal trajectory (see the black line in Fig.~(\ref{fig:system}) for three values of the longitudinal magnetic field as indicated in the legend. The right end of the lines correspond to the initial and the final states at the energy minima, while the left end of the lines corresponds to the top of the energy barrier. 
The gray shaded area marks the domain of possible $w_1$, $w_2$. 
Lables A-D indicate pairs of the eigenvalues for which the velocity diagrams illustrating the perturbation dynamics are shown in the insets. 
The background color in the insets signify whether the amplitude of the perturbation is increasing (blue), decreasing (red), or constant (gray). The damping factor $\alpha$ is $0.2$.}
\end{figure}

For zero damping, Eq.~(\ref{eq_first_order_pert}) predicts two types of dynamical trajectories for the perturbation depending on the sign of $w_1w_2$. The trajectories are elliptic, bound for $w_1w_2>0$. For the optimal switching pulse with $B_r=0$, this regime is realized in the vicinity of the energy minima for $\theta<\pi/4$ and $\theta>3\pi/4$ (see the blue regions in Fig.~\ref{fig:system}). However, the perturbation trajectories become hyperbolic, divergent for $\pi/4\le\theta\le3\pi/4$ where $w_1w_2\le 0$ (see the red region in Fig.~\ref{fig:system}).  
It is important to realize 
that for $\alpha = 0$ the trajectories are equally stable regardless of whether both $w_1$ and $w_2$ are positive or negative. Situation changes with non-zero damping: for positive $w_1$, $w_2$, 
the perturbations tend to relax toward $\vec{s}_0(t)$, while for negative $w_1$, $w_2$, the relaxation amplifies the perturbations. 
In principle, the latter case is unstable. However, this instability is expected not to significantly affect the magnetization switching if the switching time is short on the time scale of relaxation dynamics which is defined by the damping parameter $\alpha$. 
We conclude that the hyperbolic instabilities in the perturbation dynamics are the primary reason for the decoherence between the magnetization and the switching pulse. These instabilities ultimately define thermal stability of magnetization dynamics. 

The diagram in Fig.~\ref{fig2} shows evolution of $w_1$ and $w_2$ during magnetization switching. For zero $B_r$, a significant part of the switching trajectory lies in the region of unstable perturbations corresponding to the second quadrant of the diagram where the eigenvalues $w_1$ and $w_2$ have different signs. However, the values of $w_1$ and $w_2$ can be controlled by application of the longitudinal field $B_r$. In particular, the hyperbolic instabilities can be removed by shifting $w_1$ and $w_2$ either to the first ($B_r>K/\mu$) or to the third ($B_r<-K/\mu$) quadrant of the diagram in Fig.~\ref{fig2}. Therefore, the longitudinal external magnetic field can be used as a control parameter to improve thermal stability of magnetization switching. This conclusion is confirmed in the following by direct simulations of magnetization dynamics at elevated temperature ($\Delta=20$), where the switching is induced by a modified pulse $\vec{\mathcal{B}}(t)$:
\begin{equation}
    \vec{B}(t)=\vec{B}_0(t)+B_r\vec{s}_0(t).
\end{equation}



\begin{figure*}[t!]
\centering
\includegraphics[width=\textwidth]{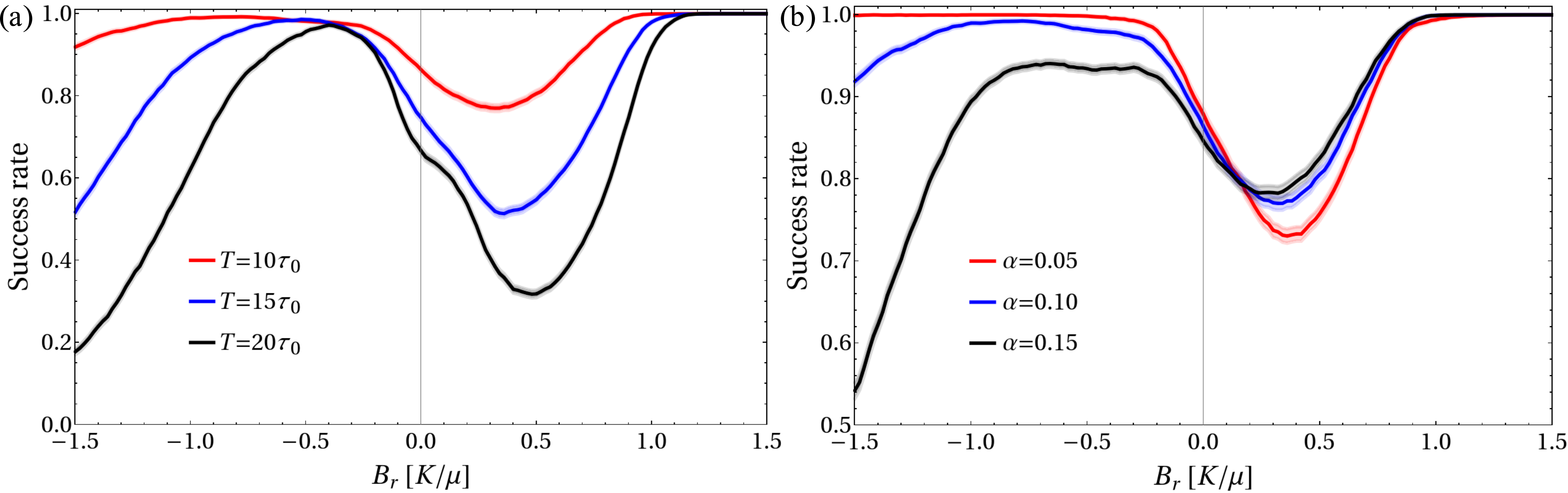}
\caption{\label{successrate_vs_Br} 
Calculated success rate of magnetization reversal as a function of the longitudinal magnetic field $B_r$ for various values of the switching time $T$ (a) and the damping parameter $\alpha$ (b). In (a), $\alpha=0.1$; 
In (b), 
$T=10\tau_0$. 
The thermal stability factor $\Delta=20$. The shaded areas around the curves indicate the statistical error.}
\end{figure*}

\textit{Effect of longitudinal magnetic field on the success rate of magnetization switching ---} Figure~\ref{successrate_vs_Br} shows calculated success rate of switching as a function of $B_r$ for various values of the switching time and damping parameter. 
As predicted, the switching success rate reaches unity for $B_r>K/\mu$ regardless of the damping factor $\alpha$ and switching time $T$. Longer switching times require stronger longitudinal field to reach a certain value of the success rate, as expected, but the threshold value of $B_r$ is not very sensitive to the damping parameter. Interestingly, the success rate as a function of the longitudinal field exhibits a minimum at $B_r\approx 0.5$ that becomes more pronounced for longer switching times. At $B_r=0.5$, the ratio between the eigenvalues becomes $w_1/w_2=-1$ at the top of the energy barrier. This corresponds to particularly unstable perturbations in the magnetization dynamics, therefore explaining the drop in the success rate of switching. 
The longer the switching time, the more time the system spends in the vicinity of the energy barrier 
\cite{kwiatkowski2021optimal}. This increases the chances of decoherence between the magnetization and the switching pulse, and lowers the switching probability. 

\begin{figure*}[t!]
\centering
\includegraphics[width=\textwidth]{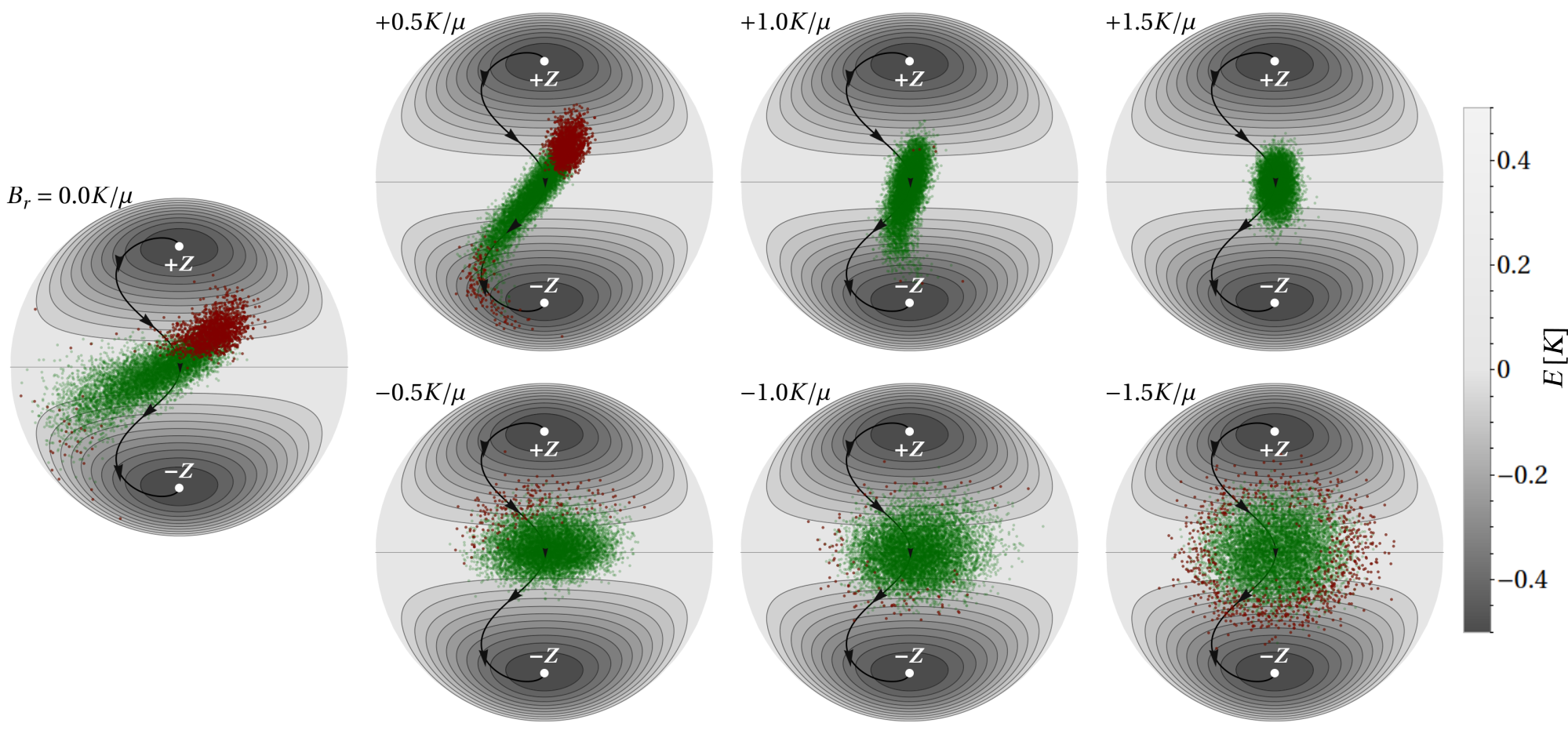}
\caption{\label{lambert_0.5T_positiveBr}
Calculated distribution of the copies of the system in the statistical ensemble at $t=T/2$ and various values of the longitudinal magnetic field, superimposed on the Lambert azimuthal projection~\cite{Snyder1987} of the energy surface of the system. The green dots correspond to the copies that will eventually reach the final state at $-Z$ (successful switching), while the red dots mark the copies that will end up at the initial state at $+Z$ (unsuccessful switching). 
The black line shows the calculated OCP for the reversal. 
The damping factor $\alpha$ is $0.1$, the thermal stability factor $\Delta$ is $20$, and the switching time $T$ is $10\tau_0$.}
\end{figure*}

Application of the longitudinal field opposite to $\vec{s}_0$ ($B_r<0$) renders both of the eigenvalues $w_1$, $w_2$ negative near the energy barrier, thus altering the hyperbolic character of the perturbation dynamics. As a result, the success rate of switching initially increases with rising $B_r$. However, further increases in $B_r$ lead to the success rate reaching a maximum value before eventually declining (see Fig.~\ref{successrate_vs_Br}).  
The drop in the success rate is a consequence of divergent dynamics due to relaxation, which becomes more prominent for larger damping parameters and longer switching times, as expected. 

\begin{figure*}[t!]
\centering
\includegraphics[width=\textwidth]{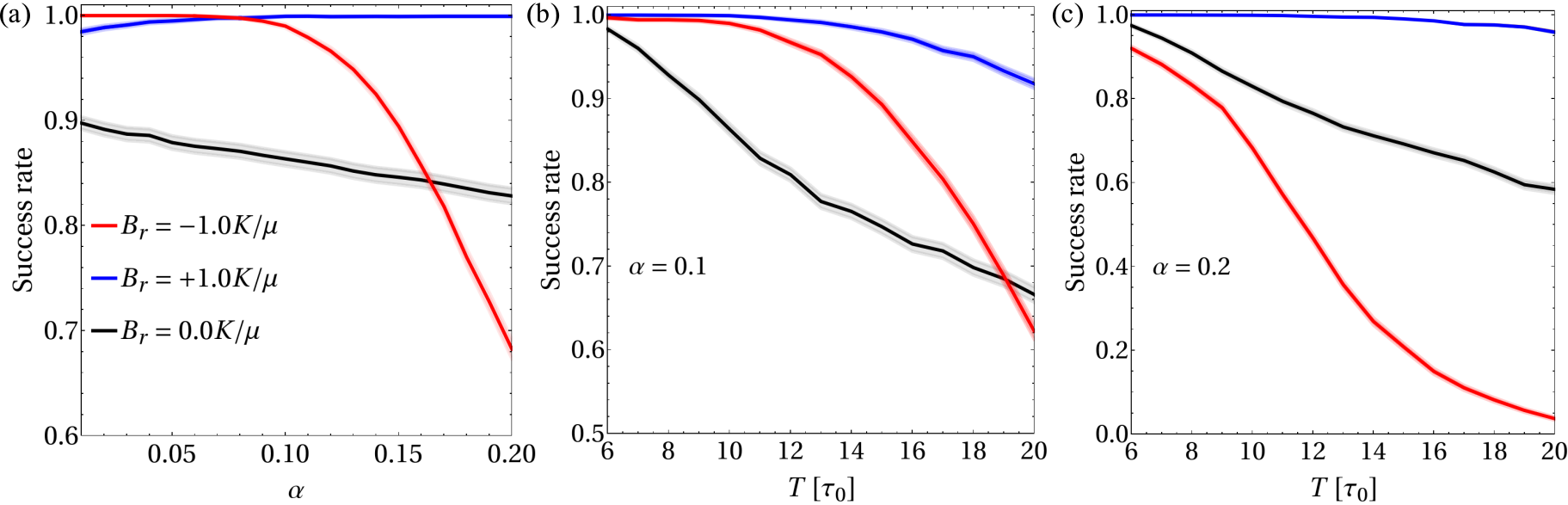}
\caption{\label{successrate_vs_alpha_and_T} 
(a) Calculated success rate of magnetization reversal as a function of damping parameter $\alpha$ for switching time $T = 10\tau_0$. (b)-(c) Calculated success rate as a function of $T$ for $\alpha=0.1$ and $\alpha=0.2$. The red, blue, and black lines correspond to the three values of the longitudinal magnetic field $B_r$ as indicated in the legend. 
The thermal stability factor $\Delta=20$. The shaded areas around the curves indicate the statistical error.}
\end{figure*}

The switching dynamics is further illustrated by Fig.~\ref{lambert_0.5T_positiveBr} showing the calculated distribution of the copies of the system in the statistical ensemble at $t=T/2$ for $\alpha=0.1$, $T=10\tau_0$, and various values of $B_r$. For the unperturbed OCP, the system is at the top of the energy barrier. Thermal fluctuations 
make the system deviate from the OCP. For zero longitudinal field, the system copies spread quite far, with those corresponding to unsuccessful switching trajectories grouped closer to the initial state. 
For $B_r = 0.5 K/\mu$, the distribution of the copies becomes more elongated -- the result of the hyperbolic character of the perturbation dynamics at the energy barrier -- and the number of the unsuccessful trajectories increases. 
As $B_r$ increases beyond $K/\mu$, a progressively tighter grouping of the copies around the OCP is observed due to the convergent dynamics of the perturbations, resulting in the switching probability approaching unity (see Fig. \ref{successrate_vs_Br}).

For negative $B_r$, the copies of the system are grouped in an ellipse around the OCP even for $B_r = -0.5K/\mu$. For stronger anti-parallel fields, the spread 
of the distribution increases due to relaxation, resulting in a decrease in the success rate of switching. 


Figure~\ref{successrate_vs_alpha_and_T} shows the calculated dependencies of the success rate on the damping constant $\alpha$ and switching time $T$ for $B_r=0$ and $B_r=\pm K/\mu$. Both cases with finite longitudinal field ensure $w_1w_2\ge0$ for the whole switching trajectory. Positive (negative) $B_r$ correspond to convergent (divergent) relaxation of the perturbation dynamics, which explains monotonic increase (decrease) of the switching probability with increasing $\alpha$.
However, for low damping and short switching times, applying the longitudinal field opposite to the magnetic moment ($B_r<0$) is more efficient than applying the longitudinal field along the magnetic moment ($B_r>0$), as it requires lower fields to achieve high success rates (see also Fig. \ref{successrate_vs_Br}). Longer switching times result in lower success rate in all considered cases, as expected. The decrease in the success rate with $T$ becomes more (less) pronounced for negative (positive) $B_r$ as damping increases, which is a result of destabilizing (stabilizing) effect of relaxation. 

\textit{Conclusions ---} In this work, we uncovered that the instability of energy-efficient protocols for magnetization reversal in nanoparticles with respect to thermal fluctuations originates from the divergent magnetization dynamics arising around the top of the energy barrier of the system. We demonstrated that these instabilities can be eliminated by applying an additional magnetic field either aligned or opposed to the magnetic moment's direction, consequently enhancing the thermal stability of magnetization switching. We examined the success rate of switching at elevated temperatures as a function of various control parameters, such as the switching time, Gilbert damping, and the magnitude of the longitudinal field. The application of a longitudinal field along the magnetic moment consistently increases the success rate of switching, provided that the field magnitude surpasses the characteristic anisotropy field. However, for shorter switching times and weaker damping, employing a smaller field opposed to the magnetic moment can also augment the success rate. 
Our results warrant a general principle for improved control of magnetization dynamics by suppressing divergent perturbations. 

\textit{Acknowledgments ---} This work was supported by the Icelandic Research Fund (Grant Nos. 217750 and 217813), the University of Iceland Research Fund (Grant No. 15673), and the Swedish Research Council (Grant No. 2020-05110).

%

\end{document}